\documentclass[a4paper,12pt]{article}
\usepackage[english]{babel}
\usepackage{amssymb,amsfonts,amsmath,mathtext,cite,enumerate,float,ulem}

\usepackage{geometry}
\newcommand{\Keywords}[1]{\par\noindent
{\small{\bf Keywords\/}: #1}}

\usepackage{setspace}

\begin{document}

\title{Correlation functions and spectral curves \\ in models of minimal gravity }
\author{O.Kruglinskaya \\ Lebedev Physics Institute \\ \small{\ttfamily{kruglinskaya@lpi.ru}}}
\date{}
\maketitle

\begin{abstract}
 The correlation functions for models of minimal gravity are discussed. An algorithm is proposed for calculations of invariant ratios from formulas of residues that can be compared with the coefficients of expansion of the partition function in Liouville theory. For (2,2K-1) models transition coefficient from basis of quasiclassical hierarchy to basis Liouville theory is obtained evidently, and the hypothesis about exact form of spectral curve has been verified.
\Keywords{generating functions, Liouville theory, integrable systems,
correlation functions.}
\end{abstract}

\section{Introduction}

It is well known that matrix models describe 2-dimensional quantum gravity with central charge $c\le 1$. In general case we should consider string loop expansion -- over the world sheets with genus more than zero, for describing theory of 2-dimensional gravity using $1/N$-expansion \cite{tHooft}.
Here we will consider only the limit $N\rightarrow \infty$, responsible for contribution of surfaces with zero genus.

The partition function of one-matix model can be presented as integral over the eigenvalues (see, for instance, \cite{1} and references therein):

\begin{equation}\label{Z}
Z=\int d \Phi \exp ^{-\frac{1}{\hbar}Tr W(\Phi)}=\frac{1}{N!}\int \prod_{i=1}^N \left( d\phi_i e^{-\frac{1}{\hbar}W(\phi_i)}\right) \Delta^2(\phi).
\end{equation}

In this expression $\Phi$ is matrix of size $N \times N$, $\hbar=\frac{t_0}{N}$ -- parameter of quasiclassical expansion (term $t_0$ remains fixed and finite at $N \rightarrow \infty, \hbar \rightarrow  0$), and $W(\Phi)$ is general determined by the formula:

\begin{equation}\label{t_k}
W(\Phi)=\sum_{k>0}t_k\Phi ^k,
\end{equation}

where $t_k$ are the sources, and $tr \Phi ^k$ -- some basis in the operator space.

The integration measure in the r.h.s. of (\ref{Z}) includes the Vandermonde determinant:

\begin{equation}
\Delta(\phi)=\prod_{i<j} (\phi_i -\phi_j).
\end{equation}

Free energy $F=\log Z$ can be represented by the expansion in series of even powers of $\hbar$, that will be equivalent to expansion in the inverse matrix size:

\begin{equation}
F\left( t_0| \{t_k \} \right)=\sum_{g=0} ^\infty \hbar ^{2g-2}F_g\left( t_0|\{ t_k \} \right), \quad N \rightarrow \infty, \hbar \rightarrow 0
\end{equation}

when $N \hbar =t_0$ is fixed.

 The main contribution in the case of large $N$ comes from only the first term $F_0$ of zero genus. Therefore we will consider only correlation functions on sphere (see \cite{Кричевер}). Below we will call $F_0=\mathcal F_{MM}$ as prepotential of matrix model by analogy with topological strings and supersymmetric gauge theories.

 Generally speaking the Seiberg-Witten system which determines prepotential $\mathcal F$ is the next set of data (see \cite{GKMMM},\cite{книга}):

\begin{itemize}
  \item family $\mathcal M$ of Riemann surfaces (complex curves) $\mathcal C$, dimension of moduli space  of this family  coincides with genus of curve $\mathcal C$
  \item meromorphic differential $dS$, whose variations over moduli are holomorphic
\end{itemize}

In terms of:

\begin{equation}\label{a1}
a_i \equiv \oint_{A_i}dS,
\end{equation}

the prepotential  $\mathcal F$ is determined from expressions:

\begin{equation}\label{a2}
\frac{\partial \mathcal F}{\partial a_i} \equiv \oint_{B_i} dS.
\end{equation}

The integration in (\ref{a1}) and (\ref{a2}) is over canonical $A$ and $B$-cycles, and formula (\ref{a2}) determines  $\mathcal F$ due to the Riemann bilinear relations.

Direct analogues of expressions (\ref{a1}),(\ref{a2}) are formulas with residues (see \cite{Кричевер},\cite{KMMM}):

\begin{equation}\label{t}
t_k=\frac{1}{k}res_{\infty} \xi ^{-k} dS, \quad k>0,
\end{equation}
\begin{equation}\label{F}
\frac{\partial \mathcal F_{MM}}{\partial t_k}=res _{\infty} \xi ^k dS, \quad k>0,
\end{equation}

  which define in the same way the prepotential of matrix model $F_0=\mathcal F_{MM}$, where $\xi$ ---  inverse local coordinate.

 The equation of spectral curve $\mathcal C$ in case of one-matrix model (\ref{Z})
\begin{equation}\label{y}
y^2=W'^2(x)+4f(x)
\end{equation}
can be obtained from loop equation at $N \rightarrow \infty$ (see \cite{Migdal}). It describes hyperelliptic curve, in terms of which free energy $F_0=\mathcal F_{MM}$ can be completely determined by (\ref{t}),(\ref{F}) with $dS=ydx$.

In equation (\ref{y}) $y=W'(x)-2G$, а $G$ is generating function (resolvent):

 \begin{equation}
G(x)=\left< Tr \frac{1}{x-\Phi} \right>=\left< \sum_{i=1}^N \frac{1}{x-\phi_i} \right>,
\end{equation}

where $f(x)$ ---  polynomial of degree on 2 less than degree of the potential $W(x)$, determined from the asymptotic $x \rightarrow \infty$.

\section{Models of minimal gravity:\\ (p,q)-critical points}

Considering matrix models as discrete analog of continuous 2-dimensional gravity, the partition function of the latter one in this approach is obtained from the partition function of matrix model as a result of non-trivial and singular double-scaling limit (see \cite{classic}). Details of this limit were studied in \cite{MMMM}, we present here only result.

Minimal conformal theories \cite{14}, interacting with Liouville gravity \cite{P81} are called (p,q)-models of minimal gravity. To calculate the partition function of (p,q)-models for zero genus we use  the same method as for calculation of prepotential of matrix model. We can write the variables $t_k$, also known as times, and free energy $\mathcal F$ using formulas where (\ref{t}),(\ref{F}), where now:

\begin{equation}
dS=YdX, \quad
\xi=X^{1/p}.
\end{equation}

For any (p,q)-point a couple of polynomials are taken:

\begin{equation}\label{полином1}
X=\lambda ^p+x_{p-1}\lambda ^{p-1}+... \, , \quad
Y=\lambda ^q+y_{q-1}\lambda ^{q-1}+... \, ,
\end{equation}

that satisfy the equation of a plane curve of the form:

\begin{equation}
Y^p=X^q+\sum C_{ij}X^iY^j.
\end{equation}

If $p=2, \, q=2K-1$, then:
\begin{equation}\label{кривая}
Y^2=P_{2K-1}(X).
\end{equation}

 which describes generally Riemann surface of genus $g=K-1$, in one case this curve becomes singular with $2K$ branch points: roots of polynomial $P_{2K-1}(X)=0$ plus $X=Y=\infty$.

According to the hypothesis proposed in \cite{11}, equation of curve (\ref{кривая}) for minimal gravity coincides with the equation on Chebyshev polynomials (in rescaled variables). Below this assumption is precised and tested for a number of models by direct calculation.

For comparison with Liouville theory ambiguity in the choice of basis needs to be taken into account, due to the fact that times have scaling dimension, that can coincide with sum of dimensions of other times in the same model

\begin{equation}
t_k \rightarrow t_k+At_{\alpha}t_{\beta}, \quad A=const,
\end{equation}
\begin{equation}\label{delta}
\Delta_k = \Delta _{\alpha}+\Delta _{\beta}, \quad \alpha,\beta<k.
\end{equation}

In such case a resonance occurs (according to the terminology of \cite{5},\cite{6}).
In formula (\ref{delta}) $\Delta_k = p+q-k$ is scaling dimension of the variable $t_k$, а $\Delta_{\alpha},\Delta_{\beta}$ --  dimensions of
$t_{\alpha}$ и $t_{\beta}$ respectively.

In $(2,2K-1)$ minimal models of gravity  cosmological constant $\mu=t_{2K-3}$ always has the dimension $\Delta _{\mu}=4$. Then resonances with cosmological constant are:

\begin{equation}\label{mm}
t_{2K-3-4i} \rightarrow t_{2K-3-4i}+ A_i\cdot \mu ^{i+1}, \quad i=1,2,... .
\end{equation}

Coefficients $A_i$  can be found from vanishing of one-point function $\frac{\partial \mathcal F}{\partial t_i}=0$ and the string equation.

 Below we describe the  scheme of calculations of transition coefficients $\{ A_i \}$ for  $(2,9)$ model. In appendix calculated conversion factors are represented for cases $K=6,7,8,9,10,11,12$.

\section{Example:(2,9) model}

For polynomials (\ref{полином1}) in this case we have:

\begin{equation}\label{polynom}
X=\lambda^2+x_0, \quad Y=\lambda^9+y_7\lambda^7+y_5\lambda^5+y_3\lambda^3+y_1\lambda.
\end{equation}

We calculate times by formulas (\ref{t}):

$$
t_1= -x_0y_1+\frac34x_0^2y_3-\frac58x_0^3y_5+\frac{35}{64}x_0^4y_7-\frac{63}{128}x_0^5,
$$
$$
t_3=\frac23y_1-x_0y_3+\frac54x_0^2y_5-\frac{35}{24}x_0^3y_7+\frac{105}{64}x_0^4,
$$
\begin{equation}\label{times}
t_5=\frac25y_3-x_0y_5+\frac74x_0^2y_7-\frac{21}{8}x_0^3,
\end{equation}
$$
t_7=\frac27y_5-x_0y_7+\frac94x_0^2, \quad t_9=\frac29y_7-x_0=0, \quad t_{11}=\frac{2}{11}.
$$

Solving latest five linear equations for $y$ and substituting solution in equation for $t_1$, we obtain string equation:

\begin{equation}
t_1+\frac32x_0t_3+\frac{63}{128}x_0^5+\frac{15}{8}x_0^2t_5+\frac{35}{16}x_0^3t_7=0.
\end{equation}

In this model only one resonance is possible:
$$
t_3 \rightarrow t_3+A_3 \mu^2, \mu=t_7.
$$

From equality to zero one-point function $\frac{\partial \mathcal F}{\partial t_3}$ and string equation we find $\mu$ and constant $A_3$:

\begin{equation}
\mu=-\frac{9}{28}x_0^2, \quad A_3=\frac{49}{36}.
\end{equation}

Free energy $\mathcal F$ of models of minimal gravity $(2,2K-1)$ can be written as:

\begin{equation}\label{f}
\mathcal F(t_i) = \mu^{K+1/2}f\left( \tau_j \right),
\end{equation}

where $f$ -- scale invariant function,
$\tau_j$ -- dimensionless ratios of times.

In this example:

\begin{equation}\label{d2}
\mathcal F(t_1,t_3,t_5, \mu)=\mu^{\frac{11}{2}}f\left( \frac{t_1}{\mu^{5/2}},\frac{t_3}{\mu^2},\frac{t_5}{\mu^{3/2}} \right), \quad \frac{\partial \mathcal F}{\partial t_1}=\mu^3 f^{(1)}, \quad \frac{\partial^2 \mathcal F}{\partial t_1^2}=\mu^{\frac12} f^{(11)}=\frac{x_0}{2}, \quad ... .
\end{equation}

and dimensionless ratios are:
\begin{equation}
\tau_1=\frac{t_1}{\mu^{5/2}}, \;
\tau_2=\frac{t_3}{\mu^2}, \;
\tau_3=\frac{t_5}{\mu^{3/2}}.
\end{equation}

  Rewriting string equation in terms of $u=f^{(11)}$:

\begin{equation}\label{string}
\tau_1+3u\tau_2+\frac{63}{4}u^5+\frac{15}{2}u^2\tau_3+\frac{35}{2}u^3=0.
\end{equation}

We expand the function $f$ in a series:

$$
f=f_0+f_1\tau_1+f_2\tau_2+f_3\tau_3+\frac12f_{11}\tau_1^2+\frac12f_{22}\tau_2^2+\frac12f_{33}\tau_3^2+f_{12}\tau_1\tau_2
$$
$$
+f_{23}\tau_2\tau_3+f_{13}\tau_1\tau_3+\frac16 f_{111} \tau_1^3+\frac16 f_{222} \tau_2^3+\frac16 f_{333} \tau_3^3+\frac12 f_{123} \tau_1\tau_2\tau_3+\frac12 f_{112}\tau_1^2\tau_2+\frac12 f_{122}\tau_2^2\tau_1+...
$$

Analitical terms $f_1$ and $f_3$ in this expansion are uninteresting, because they are not universal.

From equation (\ref{string}) one can find leading coefficients, when $\tau_2= \tau_3=0$:

\begin{equation}\label{f11}
f_{111}=-\frac{4}{105f_{11}^2 ( 3f_{11}^2+2)}, \quad f_{112}=-\frac{3}{63f_{11}^3+35f_{11}}, \quad ... .
\end{equation}

As a result we obtain dimensionless ratios, using (\ref{F}) and (\ref{d2}),(\ref{f11}), for example:

\begin{equation}\label{coeff}
\frac{f_{1111}f_{11}}{f_{111}^2}=-7, \quad \frac{f_{1111}f_{22}}{f_{112}^2}=-\frac{7}{3}.
\end{equation}

In the same way, ratios can be easily calculated for other (2,2K-1).
For instance, for model (2,11) with generating function
$\mathcal F =\mu^{13/2} \tilde{f} \left( \frac{\tilde{t_1}}{\mu^3}, \frac{\tilde{t_3}}{\mu^{5/2}}, \frac{\tilde{t_5}}{\mu^2},\frac{\tilde{t_7}}{\mu^{3/2}} \right)$:
\begin{equation}\label{coeff2}
\frac{\tilde{f}_{1111}\tilde{f}_{11}}{\tilde{f}_{111}^2}=-9, \quad \frac{\tilde{f}_{1111}\tilde{f}_{22}}{\tilde{f}_{112}^2}=-3.
\end{equation}

It is dimensionless ratios (\ref{coeff}),(\ref{coeff2})that make sense to compare with results of calculations in Liouville theory, which were obtained in recent years, using higher Liouville equations \cite{4}.

\section{Equation of spectral curve}

Now let us turn to the equation of spectral curve, that can be obtained from formulas (\ref{полином1}), where $y_i$ expressed in terms of $t_k$ using (\ref{t}). One should remember that  resonances of times $t_k$ with cosmological constant $\mu$ (formula (\ref{mm})) in a basis of Liouville theory are considered, whereupon $t_k$ assumed to be zero.

For instance, we write equation of curve for (2,9) model, contained only $\mu$ and $t_3$. To do this we express from the first equation (\ref{polynom}) $\lambda$ and substitute it in equation for $Y$. In turns all $y_i$ to be expressed through the times (\ref{times}). Making replacement we find:

$$
Y^2=\frac{1}{126}\left( X-x_0 \right) \left( x_0^4-4Xx_0^3-12X^2x_0^2+8X^3x_0+16X^4 \right)^2
$$
$$
=\frac{1}{126x_0^9}\left( X'+1 \right)\left( 2X'-1 \right)^2\left( 8X'^3-6X'-1 \right)^2.
$$

Note that r.h.s. of this equation is expressed in terms of the Chebyshev polynomial $T_9(a)$ after substitution $a=\sqrt{\frac{X'+1}{2}}$:

$$
T_9^2=(256a^9-576a^7+432a^5-120a^3+9a)^2
$$
$$
=\frac12(X'+1)(2X'-1)^2(8X'^3-6X'-1)^2.
$$

Last expression coincides with an equation of curve rescaling of variables :
\begin{equation}\label{cheb}
Y'=T_9(a), \; X'=T_2(a)=2a^2-1.
\end{equation}
and justifies Seiberg-Shih hypothesis \cite{11}.

Similarly, hypothesis can be tested for all $(2,2K-1)$ models, some of which are presented in Appendix.

\section{Conclusion}

  In this paper  algorithm  of calculation of correlators was proposed, where correlators are coefficients of expansion of generating functions in the models (2,2K-1) in the natural basis from the point of view continuous 2-dimensional gravity \cite{5}. Dimensionless ratios of this coefficients can be linked with results in Liouville theory, in which calculation are far more difficult (for example, see \cite{4}).

Also exact equations of spectral curves in Liouville basis were obtained. The hypothesis \cite{11} was tested by direct calculation, and the equations of curves were found indeed to coincide with equations of Chebyshev polynomials in rescaled variables.

Note that for general case (p,q) models of minimal gravity with arbitrary p and q an algorithm of transition to Liouville basis is still unknown.
Some some steps in this direction are, for instance, in  \cite{6},\cite{3}.

I am grateful to A.Marshakov for clarifying discussions, and A.Anokhina for valuable remarks. This work is partly supported by the Ministry of Education and Science of the Russian Federation under the contract 8498 and by RFBR grant 12-02-33095.

\newpage
\section*{Appendix}

{\small
{\begin{tabular}{|c|c|c|}
\hline
 & & \\
{\bf{(2,2K-1)}} & {\bf{Resonances}} & \\
 & \bf{with cosmological} & {\bf{String equation and equation of curve}} \\
 & \bf{constant} & \\
  & & \\
\hline
 & & \\
(2,11) & $t_{1} = \frac{729}{968}\mu^3, \quad t_{5} = \frac{81}{44}\mu^2,   $ & $t_{1}+\frac{15}{8}x_0^2t_{5}+\frac{315}{128}x_0^4t_{9}+\frac{231}{512}x_0^6+\frac{35}{16}x_0^3t_{7}+\frac{3}{2}x_0t_{3}=0$ \\
 & & \\
 & $\mu = -\frac{11}{36}x_0^2$ & $Y^2=\frac{1}{1024}(X-x_0)(x_0^5+6Xx_0^4-12X^2x_0^3-$  \\
  & & \\
  & & $-32X^3x_0^2+16X^4x_0+32X^5)^2$ \\
 & & \\
\hline
 & & \\
(2,13) & $t_{3} = \frac{6655}{4056}\mu^3, \quad t_{7} = \frac{121}{52}\mu^2,$ & $t_{1}+\frac{3}{2}x_0t_{3}+\frac{693}{256}x_0^5t_{11}+\frac{429}{1024}x_0^7+$ \\
  & & \\
 & $\mu = -\frac{13}{44}x_0^2$ & $+\frac{15}{8}x_0^2t_{5}+\frac{35}{16}x_0^3t_{7}+\frac{315}{128}x_0^4t_{9}=0$ \\
 & & \\
 &  & $Y^2=\frac{1}{4096}(X-x_0)(32X^5x_0-80X^4x_0^2-32X^3x_0^3+$ \\
  & & \\
 & & $+24X^2x_0^4+6Xx_0^5+64X^6-x_0^6)^2 $ \\
 & & \\
\hline
&  &  \\
(2,15)& $t_{1} = \frac{28561}{43200}\mu^4, \quad t_{5} = \frac{15379}{5400}\mu^3,$& $t_{1}+\frac{6435}{16384}x_0^8+\frac{3003}{1024}x_0^6t_{13}+\frac{693}{256}x_0^5t_{11}+$ \\
 & & \\
  & $t_{9} = \frac{169}{60}\mu^2, \quad \mu = -\frac{15}{52}x_0^2$& $+\frac{315}{128}x_0^4t_{9}+\frac{3}{2}x_0t_{3}+\frac{35}{16}x_0^3t_{7}+\frac{15}{8}x_0^2t_{5}=0$ \\
&  &  \\
&  & $Y^2=\frac{1}{16384}(X-x_0)(128X^7-x_0^7+64X^6x_0-192X^5x_0^2-$ \\
&  &  \\
 & & $-80X^4x_0^3+80X^3x_0^4+24X^2x_0^5-8Xx_0^6)^2$ \\
& & \\
\hline
&  &  \\
(2,17)& $t_{3} = \frac{590625}{314432}\mu^4, \quad t_{7} = \frac{10125}{2312}\mu^3,$ & $t_{1}+\frac{6435}{2048}x_0^7t_{15}+\frac{3003}{1024}x_0^6t_{13}+\frac{693}{256}x_0^5t_{11}+\frac{315}{128}x_0^4t_{9}+$ \\
& & \\
& $t_{11} = \frac{225}{68}\mu^2, \quad \mu = -\frac{17}{60}x_0^2$ & $+\frac{35}{16}x_0^3t_{7}+\frac{15}{8}x_0^2t_{5}+\frac{3}{2}x_0t_{3}+\frac{12155}{32768}x_0^9=0$ \\
 & & \\
& & $Y^2=\frac{1}{65536}(X-x_0)(256X^8+x_0^8+128X^7x_0-448X^6x_0^2-$\\
 & & \\
 & & $-192X^5x_0^3+240X^4x_0^4+80X^3x_0^5-40X^2x_0^6-8Xx_0^7)^2$\\
 & & \\
\hline
& & \\
(2,19)& $t_{1} = \frac{9938999}{16681088}\mu^5,$ & $t_{1}+\frac{46189}{131072}x_0^{10}+\frac{693}{256}x_0^5t_{11}+\frac{109395}{32768}x_0^8t_{17}+\frac{6435}{2048}x_0^7t_{15}+$ \\
 & & \\
 & $t_{5} = \frac{1753941}{438976}\mu^4,$ & $+\frac{3003}{1024}x_0^6t_{13}+\frac{15}{8}x_0^2t_{5}+\frac{3}{2}x_0t_{3}+\frac{315}{128}x_0^4t_{9}+\frac{35}{16}x_0^3t_{7}=0$ \\
 & & \\
 & $t_{9} = \frac{54043}{8664}\mu^3,$  & $Y^2=\frac{1}{262144}(X-x_0)(256X^8x_0-1024X^7x_0^2-448X^6x_0^3+$ \\
   & & \\
  & $t_{13} = \frac{289}{76}\mu^2,$ &$+672X^5x_0^4+240X^4x_0^5-160X^3x_0^6-$ \\
     & & \\
  & $\mu = -\frac{19}{68}x_0^2$& $-40X^2x_0^7+10Xx_0^8+512X^9+x_0^9)^2$ \\
   & & \\
 \hline
\end{tabular}}
}

\newpage
{\small
{\begin{tabular}{|c|c|c|}
\hline
 & & \\
 (2,21) & $t_3=\frac{2476099}{1185408}\mu^5, $ & $t_{1}+\frac{88179}{262144}x_0^{11}+\frac{109395}{32768}x_0^8t_{17}+\frac{230945}{65536}x_0^9t_{19}+$ \\
    & & \\
   & $t_7=\frac{1433531}{197568}\mu^4,$ & $+\frac{35}{16}x_0^3t_{7}+\frac{315}{128}x_0^4t_{9}+\frac{6435}{2048}x_0^7t_{15}+$ \\
    & & \\
 & $t_{11}=\frac{89167}{10584}\mu^3,  $ &$+\frac{693}{256}x_0^5t_{11}+\frac{3003}{1024}x_0^6t_{13}+\frac{15}{8}x_0^2t_{5}+\frac{3}{2}x_0t_{3}=0$
   \\
      & & \\
         &  $t_{15}= \frac{361}{84} \mu^2,$& $Y^2=\frac{1}{1048576}(X-x_0)(512X^9x_0-2304X^8x_0^2-$ \\
    & & \\
       & $\mu=-\frac{21}{76}x_0^2$& $-1024X^7x_0^3+1792X^6x_0^4+672X^5x_0^5-560X^4x_0^6-$ \\
     & & \\
       & & $-160X^3x_0^7+60X^2x_0^8+10Xx_0^9+1024X^{10}-x_0^{10})^2$ \\
   & & \\
   \hline
      & & \\
 (2,23) & $t_1=\frac{1801088541}{3295407616}\mu^6,$ &$t_{1}+\frac{676039}{2097152}x_0^{12}+\frac{3003}{1024}x_0^6t_{13}+\frac{6435}{2048}x_0^7t_{15}+$\\
    & & \\
   & $t_5=\frac{943427331}{179098240}\mu^5,$ & $+\frac{230945}{65536}x_0^9t_{19}+\frac{969969}{262144}x_0^{10}t_{21}+\frac{109395}{32768}x_0^8t_{17}+$ \\
    & & \\
     & $t_9=\frac{9270261}{778688}\mu^4,$ & $+\frac{693}{256}x_0^5t_{11}+\frac{315}{128}x_0^4t_{9}+\frac{35}{16}x_0^3t_{7}+\frac{15}{8}x_0^2t_{5}+\frac{3}{2}x_0t_{3}=0$ \\
   & & \\
   & $t_{13}=\frac{46305}{4232}\mu^3,$ & $Y^2=\frac{1}{4194304}(X-x_0)(-x_0^{11}+2048X^{11}+1024x_0X^{10}-$ \\
   & & \\
   & $t_{17}=\frac{441}{92}\mu^2, $ & $-5120x_0^2X^9-2304x_0^3X^8+4608x_0^4X^7+1792x_0^5X^6-$ \\
    & & \\
     &$\mu=-\frac{23}{84}x_0^2$ & $-1792x_0^6X^5-560x_0^7X^4+280x_0^8X^3+60x_0^9X^2-12x^{10}_0X)^2$\\
    & & \\
 \hline
\end{tabular}}
}

\end{document}